# Predicting Risk of Developing Diabetic Retinopathy using Deep Learning

Ashish Bora, MS[1]
Siva Balasubramanian, MD PhD[2]
Boris Babenko, PhD[1]
Sunny Virmani, MS[1]
Subhashini Venugopalan, PhD[3]
Akinori Mitani, MD PhD[1]
Guilherme de Oliveira Marinho, BS[4]
Jorge Cuadros, OD PhD[5]
Paisan Ruamviboonsuk, MD[6]
Greg S Corrado, PhD[1]
Lily Peng, MD PhD[1]
Dale R Webster, PhD[1]
Avinash V Varadarajan, MS[1]
Naama Hammel, MD[1]
Yun Liu, PhD[1*]
Pinal Bavishi, BE[1*]

Affiliations:
[1]Google Health, Google LLC, 1600 Amphitheatre Pkwy, Mountain View, CA 94043, USA
[2]Work done at Google Health via Advanced Clinical, Deerfield, IL, USA
[3]Google Research, Google LLC, 1600 Amphitheatre Pkwy, Mountain View, CA 94043, USA
[4]Work done at Google Health via Adecco Staffing, Santa Clara, CA, USA
[5]EyePACS LLC, Santa Cruz, CA, USA
[6]Department of Ophthalmology, Rajavithi Hospital, College of Medicine, Rangsit University, Bangkok, Thailand

*Equal contribution

Corresponding Author:
Naama Hammel, MD
Google Health, Google LLC
1600 Amphitheatre Pkwy, Mountain View, CA 94043, USA
nhammel@google.com

Diabetic retinopathy (DR) screening is instrumental in preventing blindness, but faces a scaling challenge as the number of diabetic patients rises. Risk stratification for the development of DR may help optimize screening intervals to reduce costs while improving vision-related outcomes. We created and validated two versions of a deep learning system (DLS) to predict the development of mild-or-worse ("Mild+") DR in diabetic patients undergoing DR screening. The two versions used either three-fields or a single field of color fundus photographs (CFPs) as input. The training set was derived from 575,431 eyes, of which 28,899 had known 2-year outcome, and the remaining were used to augment the training process via multi-task learning. Validation was performed on both an internal validation set (set A; 7,976 eyes; 3,678 with known outcome) and an external validation set (set B; 4,762 eyes; 2,345 with known outcome). For predicting 2-year development of DR, the 3-field DLS had an area under the receiver operating characteristic curve (AUC) of 0.79 (95%CI, 0.78-0.81) on validation set A. On validation set B (which contained only a single field), the 1-field DLS's AUC was 0.70 (95%CI, 0.67-0.74). The DLS was prognostic even after adjusting for available risk factors ($p<0.001$). When added to the risk factors, the 3-field DLS improved the AUC from 0.72 (95%CI, 0.68-0.76) to 0.81 (95%CI, 0.77-0.84) in validation set A, and the 1-field DLS improved the AUC from 0.62 (95%CI, 0.58-0.66) to 0.71 (95%CI, 0.68-0.75) in validation set B. The DLSs in this study identified prognostic information for DR development from CFPs. This information is independent of and more informative than the available risk factors.

**Funding**
Google LLC, Mountain View, California, funded this study and had a role in its approval for publication.

# Research in context

**Evidence before this study**
A review of the current literature on predicting the progression risk of diabetic retinopathy (DR) was conducted using Google Scholar, PubMed, Scopus, MEDLINE and Web of Science. The search resulted in studies published in English until 15 June, 2020 using the keywords "progression of diabetic retinopathy", "predicting development of diabetic retinopathy", "predicting incident diabetic retinopathy", "risk factors for diabetic retinopathy", "diabetic retinopathy screening programs", "automated detection of diabetic retinopathy using fundus photographs", "artificial intelligence", "machine learning" and "deep learning". Although few studies have investigated the risk of DR progression on fundus photographs using deep learning, to our knowledge no study has reported automated risk stratification of developing DR from no DR on fundus photographs or this prediction in combination with known risk factors.

**Added value of this study**
In this study, we created a deep learning system (DLS) that uses color fundus photographs (CFPs) to predict the development of mild-or-worse ("Mild+") DR within two years, for diabetic patients without DR. The DLS was created using a large retrospective longitudinal dataset collected in a DR screening program. Significant improvement in risk stratification was observed on two independent datasets from different countries and ethnicities. The use of an automated risk stratification tool could help scale DR screening as the number of patients with diabetes continues to grow.

**Implications of all the available evidence**
Our risk stratification tool could be used by clinicians to optimize DR screening intervals. Diabetic patients identified by our algorithm as high risk could be followed up more closely to improve visual outcomes, while lower risk patients could be followed up less frequently to reduce the burden of screening. More effective screening strategies are particularly important given the recent challenges introduced by the COVID-19 pandemic. The risk prediction, in concert with personalized patient education could motivate patients to undertake lifestyle modifications or increase their adherence to therapy for improved blood sugar control. Finally, our tool could aid further research into diabetes management and treatment. Prospective studies are required to validate the DLS as a risk predictor for DR development and its effectiveness in real world screening programs and clinical settings.

## Introduction

Diabetic retinopathy (DR) is the leading cause of irreversible blindness among adults aged 20-74 years globally.[1] Because treatments are available to slow, halt, or effectively reverse progression during the early stages of DR, regular screening is key to preventing blindness. However, since the total number of diabetic patients is expected to increase from 415 million in 2015 to 642 million in 2040,[2] the burden of screening and subsequent follow-up is a significant challenge. The efficiency of DR screening programs may be improved using risk stratification tools that can help personalize screening frequencies based on the likelihood of development or progression of DR.[3] In this work, we created a deep learning system (DLS) that uses color fundus photographs (CFPs) to predict the risk of developing DR.

DR is a consequence of injury to the retinal microvasculature from diabetes. Most diabetic patients develop DR during the first two decades of having diabetes.[4] Without suitable intervention, DR can develop and eventually progress from mild or moderate stages to vision-threatening diabetic retinopathy (VTDR: severe nonproliferative DR, proliferative DR, and/or diabetic macular edema).[5] With the goal of early detection, DR screening is widely recommended by major organizations such as the American Diabetes Association (ADA),[6] International Council of Ophthalmology (ICO)[7] and American Academy of Ophthalmology (AAO).[8] The recommended screening frequencies vary based on the local resource availability and the stage of DR. For no or mild DR, the recommended frequency ranges from every 12 to every 24 months.

Signs of retinal microvascular changes caused by diabetes are visible in CFPs, and thus they are routinely used to assess the stage of DR. In addition to DR stage, the risk of development and progression is influenced by systemic risk factors. Modifiable risk factors include hyperglycemia, hypertension, dyslipidemia and obesity, smoking, anemia, pregnancy, low health literacy, inadequate healthcare access, and poor adherence to therapy.[9,10] Ethnicity, family history or genetics, age at onset, type of diabetes, and duration of diabetes are non-modifiable risk factors for the development and progression of DR.[9]

In this work, we focus on using CFPs, in addition to several known risk factors, to improve stratification of risk of developing DR. Notably, this is a challenging task, since microvascular changes are not known to be detectable in CFPs prior to development of DR. Consequently, current CFP-based grading systems, such as the Early Treatment Diabetic Retinopathy Study (ETDRS),[11] International Clinical Diabetic Retinopathy (ICDR),[12] and The Royal college of Ophthalmologists Diabetic Retinopathy guidelines[13] do not further risk stratify such cases. However, since the majority of patients in a DR screening program show no signs of DR (>85% in both datasets in our study), and typically only a small fraction of these patients require retinal intervention within 2 years,[14] risk stratification of this subgroup is an important problem. We hypothesized that for patients who show no signs of DR, a DLS using CFPs may be able to provide such stratification.

Using a longitudinal dataset of CFPs, we created a DLS to predict whether eyes will develop mild-or-worse ("Mild+") DR within the following two years. In two validation datasets, we show significant improvements in risk stratification both using CFPs alone, and when combined with available risk factors as compared to using risk factors only. Additionally, we show that the learned risk stratification generalizes across outcomes at different time points as well as when evaluated against more severe stages of DR. We believe this system could aid in clinical-decision making such as optimizing screening intervals, target lifestyle modifications or counseling, and enable further research in diabetes management and treatment.

## Methods

### Datasets

We used de-identified CFPs from two datasets (Table 1). The first dataset comprised 362,283 patients with a total of 430,917 retinal examination visits from EyePACS LLC, a teleretinal DR screening service in the U.S.. The standard imaging protocol was to capture three 45° fields of CFPs (nasal, primary, and temporal) per eye. Different devices were used at the screening sites including Canon (CR1 and CR2), Topcon (NW200 and NW400), Zeiss Visucam, Optovue iCam, and Centervue DRS. Visits lacking any of the three fields were excluded. The average follow-up amongst patients with two or more visits was 26.7 months (median: 21.4, interquartile range: 13.8-33.9, range: 0-126). The EyePACS dataset was randomly split into development and our first validation set (validation set A) in an 8:2 ratio while ensuring no patients overlapped. The development set was used to train and tune a DLS to predict the development of DR amongst eyes without DR at their first visit (see details in "Algorithm development"). The DLS was then validated on validation set A, which contained 7,976 eyes after selecting a random eye per patient and restricting to eyes with more than two visits where the first visit had no DR. The histogram of inter-visit time is plotted in Supplementary Figure S1A and the HbA1c distribution is visualized in Supplementary Figure S2.

The second dataset was from Thailand and contained 6,791 patients.[15] Patients were randomly identified from the Thai national diabetic patients registry, representing hospitals or health centers in each of the 13 health regions in Thailand. The standard imaging protocol involved taking 45° primary field CFPs. The dataset contained images captured using various devices: Topcon TRC-NW8, Nidek (AFC-210 and AFC-230), KOWA (Nonmyd α-DIII 8300, Nonmyd 7, VX-10α, Nonmyd α-DIII, Nonmyd WX, VX-20). The histogram of inter-visit time is plotted in Supplementary Figure S1B. Most patients were followed up after approximately 2 years (average: 21.4 months, median: 23, interquartile range: 19-24, range: 2-35). One random eye per patient was selected, and both visits for those eyes were graded (see details in "Grading procedure"). When considering only those with two gradable visits and no DR at the first visit, 4,762 eyes remained. This dataset was used as a second, external validation set (validation set B). The HbA1c distribution for this dataset is visualized in Supplementary Figure S2.

All images and metadata were de-identified according to the Health Insurance Portability and Accountability Act (HIPAA) Safe Harbor provision before transfer to the study investigators. Ethics review and institutional review board exemption was obtained from Advarra Institutional Review Board.

### Endpoint definition

This work dealt with predicting development of DR, defined as whether an eye without DR at the initial visit was graded as having Mild+ DR within 2 years. Because DR can develop in between two visits, a Mild+ grade given during a visit implies that the disease progression happened between the prior and the current visit. To account for this and real-world scheduling logistics for in-person DR screening visits, we used a "buffer" of 1 month (28 days) for this binary endpoint definition. Specifically, we considered an eye as positive for our endpoint if a visit corresponding to Mild+ DR happened within 2 years after the initial visit or during the month following the 2-year mark. For eyes that did not meet this criterion, we required that the eye had a no-DR visit after 2 years, or during the month prior to the 2-year mark to be considered negative. Eyes that did not satisfy either of these criteria were excluded for evaluations involving this binary endpoint. However, eyes without a known 2-year Mild+ DR endpoint were considered for the survival analysis (see "Algorithm evaluation" below).

### Grading procedure

Images from EyePACS (development set and validation set A) were graded for DR lesions by EyePACS-certified graders based on a modified ETDRS grading protocol.[16] Graders graded each eye based on all three fields (nasal, primary, and temporal). The absence, or presence and severity of individual lesions were mapped to 5 DR levels (no DR, mild, moderate, severe, and proliferative DR) and diabetic macular edema (DME) (Supplementary Table S1). Each image from the Thailand dataset (validation set B) was graded by one retina specialist from a pool of 11 specialists (average years of experience: 4 years; range: 1-9) using a similar grading protocol (Supplementary Table S2). Images that were considered ungradable were excluded from the study.

### Algorithm development

A DLS based on the Inception-v3[18,19] architecture was trained on the train set (⅞-th of the development dataset) and tuned using the tuning set (⅛-th of the development set). Within the training set, only 25,211 out of 503,527 eyes had known 2-year Mild+ DR outcomes (Table 1). To make use of all of these data, we leveraged the fact that all eyes were labeled with the presence or absence of lesions and corresponding DR grade using multi-task learning. Specifically, we tasked the DLS with simultaneously learning to grade DR and DME, detect the presence of individual lesions, and predict other outcomes such as moderate-or-worse DR and VTDR at multiple future time points. More details and specific hyperparameters are reported in

Supplementary Table S3. Alternative modeling approaches are described in Supplementary Methods and Supplementary Table S4.

We developed two versions of the DLS, one that takes as input the primary field (indicated as "1-field"), and another that takes three fields as input (nasal, primary, and temporal; indicated as "3-field"). We present the results for both versions of the DLS for the EyePACS validation set (that contains all 3 fields). However, because the Thailand dataset contained only the primary field, we evaluated only the 1-field version on this dataset. Both versions of the DLS take as input CFPs of size 587x587 pixels and output a number between 0 and 1 representing the likelihood of an eye developing DR within 2 years.

Rescaling of predicted likelihoods was found (post-hoc) to be necessary to calibrate predicted probabilities in the external validation set B. Specifically, we applied the DLS to 5% of the eyes in validation set B, and empirically determined a constant factor for linearly rescaling the predicted likelihood (between 0 and 1) such that the average rescaled likelihood matched the incidence of DR within 2 years. This resulting scaling factor (0.65) was then used on the full validation set B to produce calibration plots.

### Algorithm evaluation

We evaluated the DLS's ability to predict development of DR in multiple ways. First, we evaluated the discriminatory performance using the area under the receiver operating characteristic (area under ROC; AUC). Next, we evaluated the calibration of the model by plotting the observed event rate against the predicted event rate, based on deciles of risk predicted by the DLS. Additionally, we evaluated the positive and negative predictive values for the DLS predictions at all percentiles of DLS prediction.

To further evaluate the DLS with respect to outcomes that occurred beyond 2 years, we conducted survival analysis with Kaplan-Meier curves, log-rank tests, and Cox proportional hazards regression models. For the Kaplan-Meier curves and log-rank tests, thresholds for high-risk and low-risk were based on the upper and lower quartiles of DLS prediction in the tuning set.

### Comparison and integration with risk factors

Finally, to compare prognostication by the DLS to that by risk factors at the time of the first visit, we trained univariable and multivariable logistic regression models on the development set, and evaluated them on the validation sets. Different risk factors were available in each dataset, and each experiment included only the patients who had those risk factor(s) available. For EyePACS, HbA1c, self-reported diabetic control, years with diabetes, and insulin use were available. Diabetic control was reported as poor, fair, moderate, good, or excellent. Insulin use was defined as either patient self-reported, or when present in the electronic medical record. For the Thailand dataset, HbA1c and hypertension status were available. Categorical variables were encoded as dummy variables. The risk factors were also processed to handle extreme outlier values (Supplementary Methods).

### Statistical analysis

Survival analysis was conducted in Python using the lifelines library v0.24.6. Confidence intervals for proportions (sensitivity, specificity, positive/negative predictive values) were calculated using the Clopper-Pearson method based on the Beta distribution (as implemented in Python's statsmodels library, v0.10.1), for the AUC using the DeLong method, and for the Kaplan-Meier plots using the Greenwood Exponential formula. Statistical significance was evaluated using the likelihood ratio test for the Cox models, and the logrank test for the Kaplan-Meier analysis.

### Model Explanation

To understand how the DLS was able to predict the development of DR from CFPs, we used two strategies. First, we used several saliency techniques (Integrated Gradients, Guided Backprop and Blur Integrated Gradients)[20–22] that highlight the pixels with the greatest impact on the DLS's prediction for *each individual image*. For example, this enables us to identify areas of the CFP the DLS might have been using to make a prediction for each case that progressed, but for which there were no signs of DR. The second strategy we adopted was an "input ablation" approach: systematically including or excluding certain fields or regions of each image by replacing them with a mask during training as well in evaluation. For the region-based analysis, the outer rim and top and bottom parts of the image were cropped slightly to ensure consistent contours between images from different devices, and the processed image was split into 5 regions with equal area (Figure 5C). The consequent drop in performance allows us to identify retinal areas that are most important in prognostication generally *across the whole dataset*.

## Results

In this study we used two datasets; the DLS was developed using a subset of the EyePACS dataset (train set, n=503,527 eyes; tune set, n=71,904 eyes), and evaluated on an independent subset of EyePACS (validation set A, n=7,976 eyes) and the Thailand dataset (validation set B, n=4,762 eyes). The baseline characteristics, demographics and distribution of the DR grades from the datasets used are shown in Table 1. Amongst eyes with known 2-year Mild+ DR outcomes in the two validation datasets (n=3,678 and 2,345, respectively), 15-19% of the eyes without DR developed some level of DR within 2 years.

The 3-field DLS's AUC for predicting development of DR in validation set A was 0.79 (95%CI: 0.77-0.81). The 1-field DLS's AUC was 0.78 (95%CI: 0.76-0.80) in validation set A and 0.70 (95%CI: 0.67-0.74) in validation set B. These results are illustrated in Supplementary Figure S3A-C, while corresponding results for a subset of both validation sets that had available risk factors were nearly identical and presented in Figure 1A-C. The comparison with risk factors will be detailed in a subsequent section. Both versions of the DLS were well-calibrated for validation set A (Figures 1D-E). In validation set B, however, consistent with a substantially lower incidence of DR (15% vs 19%), the 1-field DLS over-estimated the rate of developing DR. This overestimation was resolved after a simple scaling using 5% of validation set B (Figure 1F). This calibration does not affect the AUC analysis.

Positive predictive values (PPV) and negative predictive values (NPV) as a function of the risk threshold are illustrated in Figure 2. Because the incidence rates were below 20%, the NPVs were generally higher than 80%, exceeding 95% for the lowest-risk group. The PPVs increased sharply in the high-risk group, with the predicted risk of developing DR in the highest-risk decile of patients ranging from 40% to 60%.

Results from univariable analyses comparing the DLS to known risk factors, and multivariable analysis evaluating if the DLS could improve prognostication when added to risk factors are presented in Table 2A and Figure 1A-B for validation set A. The most predictive risk factor in univariable analysis was HbA1c (AUC of 0.68). The remaining risk factors had AUCs ranging from 0.59 to 0.64. The combination of all risk factors using multivariable logistic regression achieved an AUC of 0.72. Upon adding the prediction of the 3-field DLS as another variable to the logistic regression model, the AUC increased from 0.72 to 0.81. The effect of instead adding the prediction of the 1-field DLS was similar albeit with slightly smaller improvements, with the AUC increasing to 0.79. In the same cohorts, the DLS alone achieved consistent AUCs of about 0.79 using 3-fields and 0.77-0.78 using 1-field. The results of several weakly-prognostic risk factors are reported in Supplementary Table S5. In validation set B, the only overlapping risk factor with the development sets was HbA1c. Thus, we next compared HbA1c alone to a model that also includes the DLS (i.e., the model from the first row of Table 2A). The addition of the DLS improved the AUC from 0.62 to 0.71 in validation set B (Table 2B and Figure 1C). In contrast to adding the DLS to risk factors, the difference between using the DLS in isolation and adding risk factors to the DLS was much more modest at 0.01-0.02 absolute improvement in both validation sets.

Next, we used Kaplan-Meier analysis to examine the association of the DLS with development of DR over time. First, we defined high-risk, medium-risk, and low-risk groups using the thresholds for upper and lower quartiles of DLS prediction in the tuning dataset. Kaplan-Meier plots illustrating rates of development of DR are presented in Figure 3A for the 3-field DLS on validation set A, and Figures 3B-C for the 1-field DLS on the two validation sets. A clean separation between the low and high risk groups' rates of development of DR was observed in these evaluations (p<0.001). Corresponding analyses for how the same risk groups progressed to moderate-or-worse DR and VTDR are presented in Figures 3D-I. We observed that the risk groups remained separated even when progression to more advanced stages of DR are considered (p<0.001).

Cox regression analysis is presented in Supplementary Tables S6A-B. The DLS predictions were significantly associated (p<0.001) with development of DR both in univariable analysis and after adjusting for HbA1c, self-reported diabetic control, years with diabetes, and insulin use in validation set A. Similar results were observed in validation set B, with multivariable adjustment for HbA1c and self-reported hypertension.

Next, we evaluated if there was a temporal relationship between the predicted risk of developing DR and the time-to-DR (Supplementary Figure S4). At 4-8 years before DR developed, the baseline median predicted DR risk remained constant at approximately 10%. The median predicted risk increased to approximately 20% 3 years prior to DR, to approximately 25% 2 years before DR, and to approximately 40% 1 year before DR developed.

To better understand the CFP features that our system had identified as being prognostic for developing DR, we employed three explanation techniques: Integrated Gradients, Guided Backprop, and BlurIG.[20–22] We present results for four cases using the Integrated Gradients method in Figure 4 and defer the full set of results to Supplementary Figure S5. For cases without signs of DR at baseline but that later developed DR, we observed both cases where the heatmaps highlighted areas which eventually developed lesions (Figure 4A-B), and cases where the heatmap highlights did not show lesions at the follow-up visit (Figure 4C). For the case in Figure 4D that already had subtle microaneurysms, the heatmaps highlighted the existing lesions. Figure 4E shows a case that did not develop DR in 6 years of followup.

Finally, we present results from input ablation experiments (Methods). For the first set of experiments, we considered the 3-field DLS, and ablated either one or two fields of input (Figure 5A-B). Among single-field setups (ie, when removing two fields), the highest AUC was achieved when retaining the primary field. Among the two-field setups (i.e., when removing one field), the highest AUC was achieved when the fields covered the widest field of view (nasal and temporal). In a second set of experiments, we considered the 1-field DLS and split the primary field into 5 regions (see Methods) for inclusion or exclusion (Figure 5C-D). The central region alone resulted in the highest AUC, while the inferior region alone resulted in the lowest AUC.

## Discussion

In this study we created a DLS to predict development of DR within 2 years, and validated it on two datasets. The first, internal validation set contained images from predominantly (59%) Hispanic patients, while the second, external validation set were from Thai patients. On both datasets, the DLS demonstrated good performance both in isolation, and when adjusted for available risk factors. When combined with several risk factors including HbA1c, the performance improved from using the risk factors alone. The Kaplan-Meier analysis added further supportive evidence that the prognostication generalized to different time points and that the model's predictions were also associated with progression beyond mild DR; to moderate DR and VTDR as well.

The model was well calibrated for validation set A, but overestimated DR incidence in validation set B. This overestimation may be associated with the lower incidence rate in validation dataset B (15% vs 19%), which is in turn consistent with a lower proportion of patients with high HbA1c (Supplementary Figure S2). Differences in grading protocols (Supplementary Table S1 & S2) may also have contributed to this. Specifically, the presence of hemorrhages or hard exudates without microaneurysms were considered moderate DR in validation set A, but no DR in validation set B. This affects our inclusion criteria and the endpoint. Moreover, validation set A was labeled by trained graders, whereas validation set B was labeled by retina specialists for the purposes of this study. The calibration difference was resolved when 5% of validation set B was used to recalibrate the model. In real use cases, retrospective data may suffice for recalibration, though this will need to be studied in future work.

In recent years, deep learning applied to CFPs or OCTs has shown great promise for many diagnostic tasks.[23–25] In ophthalmology, deep learning has been used to automatically detect or grade DR.[26–31] The first approval from the U.S. Food and Drug Administration (FDA) for an autonomous artificial intelligence system for DR detection was granted in 2018.[32] Going a step further, this approach has tackled tasks not known to be possible even for human experts, suggesting that deep learning models can potentially detect novel signs of disease. For instance, models using CFPs have been shown capable of detecting anemia,[33] predicting cardiovascular risk factors,[34] and determining DME and glaucoma-related damage that otherwise needed optical coherence tomography (OCT) or additional tests for a diagnosis,[17,35,36] predict the progression of DR,[37] progression of age-related macular degeneration,[38–40] and development of glaucoma.[41]

There are also several examples of algorithms for stratifying risk of DR. Estil *et al.* used an algorithm to reduce the frequency of screening visits based on individual risk factors for DR progression.[42] Cunha-Vaz *et al.*,[43] found that in eyes with mild non-proliferative DR, increased microaneurysm turnover rate and central macular thickness were correlated with an increased risk of progression to DME.[43] Using deep learning, Arcadu *et al.*[37] predicted DR progression, defined as worsening of 2 or more steps on the ETDRS scale at 6, 12 and 24 months, with AUCs of 0.68, 0.79 and 0.77, respectively. The limitations of the study were lack of adjusted analysis for risk factors, lack of external validation set, limited study size (n≈530) and the consequent use of cross-validation, and a restrictive inclusion criteria (patients from two clinical trials). A multivariable model developed using multifocal electroretinogram achieved 86% sensitivity and 84% specificity for predicting new retinopathy development at specific locations in the retina over one and two years.[44] A study by Rogers *et al.*,[45] found that retinal arteriolar dilation increased the risk of incident DR over 5 years in an Australian cohort suggesting retinal arteriolar dilatation could be a preclinical biomarker for DR.

Our study improves upon prior work in several ways. Firstly, we consider the challenging task of stratifying patients by their risk of developing DR using CFPs and risk factors, which are available in most screening settings. This directly tackles the problem of optimizing screening intervals by stratifying the largest group in screening: those without any DR. Second, our system retained high prognostic value after adjusting for several risk factors which suggests the development of independent imaging-based biomarkers is a fertile area of further research. Thirdly, we validated our system on two separate validation datasets from two continents. Both datasets represented their region's diabetic screening patient population; and our system retained substantial predictive value despite differences in HbA1c levels (Supplementary Figure S2), average incidence rates (Table 1), and grading protocols (Supplementary Tables S1 and S2). In addition, our development sets spanning multiple years of follow-up are orders of magnitude larger than prior work. Finally, the risk groups constructed based on our system retain significant separation even when evaluated against different endpoints (moderate-or-worse and VTDR plots in Figure 3D-I).

Moreover, we conducted several analyses to better understand the features in the CFPs that are predictive of future development of DR. First, our model predicts a very high likelihood of developing DR when there already exists subtle microaneurysms that were missed by the initial grader. While this signifies cases that were meant to be excluded from the study, subtle microaneurysms can be missed in real settings. As such, it is comforting that such cases are flagged with high certainty by the DLS, which could prompt graders to take a more careful look at the eye. Second, the saliency (heatmap) analysis in Figure 4 and Supplementary Figure S5 suggests that the DLS sometimes "focuses on" anatomic regions that eventually develop DR. This suggests that there may be subtle signs that are not visually apparent, and this phenomenon merits study in future work. Third, in terms of the field, the primary field was found to be most important. At the same time, the best combination of two fields was the nasal and temporal (excluding the primary field). These observations can be reconciled by considering that (1) the combination of the nasal and temporal fields encompasses the primary field, and (2) these two fields when combined provide the most expansive view of the retina. Finally, within the primary field, we observed that the superior and inferior regions seemed to be the least important by virtue of having the least impact when excluded. Conversely, the macular region had the greatest impact when removed and was also the most prognostic in isolation. Thus it seems important for the DLS to have both information about the macula specifically, and information from the periphery of the retina to predict the development of DR.

In terms of utility, one application of the ability to predict the development of DR can be in optimizing screening intervals. The availability of effective treatments such as laser photocoagulation and intravitreal injections of anti-vascular endothelial growth factor[46–48] have elevated the importance of early detection via screening. Traditionally, DR screening was performed by a clinician using direct or indirect ophthalmoscopy or slit lamp biomicroscopy. However, fundus photography and its review by a remote clinician is becoming more widespread because of its cost-effectiveness and acceptable sensitivity and specificity rates.[49] Multiple DR screening guidelines recommend screening intervals based on assessment of CFPs.[7,8,50] However, these guidelines provide a range of values for the screening interval (e.g. 12-24 months for patients with no apparent DR) to account for clinical risk factors, resource availability, and other socioeconomic factors. Our tool can provide a more accurate personalized risk assessment to help determine optimal screening intervals; high-risk patients can be followed frequently to ensure early detection, while low-risk patients could be followed less frequently to reduce screening burden shared by patients, clinicians, and the healthcare system. Such strategies based on the risk assessments, and corresponding interventions will need to be studied in future work.

Other potential use cases of predicting the development of DR revolve around targeted interventions. For example, high-risk patients could be preferentially selected for more intensive lifestyle modifications or counseling.[51,52] These patients may also be better suited for stricter pharmacologic control of blood sugar, and could be suitable candidates for participation in relevant clinical trials. Patients sometimes miss their screening visits, and in such cases, our tool can be used to alert high-risk patients (e.g. via phone calls or emails) to improve screening compliance and consequently visual outcomes. Finally, the ability of the DLS to predict the development of DR that occurs beyonds 2 years (Figure 3) suggests the tool may also be useful in long-term forecasting of population-level DR prevalence for public health planning.

## Limitations

Our study has several limitations. First, several known risk factors such as blood pressure were not available for both datasets, and so we could not compare the DLS to these risk factors, or adjust for them in multivariable analysis. Some risk factors such as HbA1c were also potentially from an earlier blood test, which adds uncertainty to the HbA1c at the time of DR screening. Second, as demonstrated in Figure 4, grading variability exists, especially for subtle findings such as microaneurysms. As such, a subset of our cohort already had mild DR at baseline. Though it is semantically correct that our model predicted these images as "will develop DR" with high certainty, users of a tool such as this should be aware that eyes with unreasonably high predicted risk may have already developed DR. Potential solutions to this include the use of automated DR grading algorithms in parallel with a risk stratification tool, and could be investigated in future work. Next, since our grades were based on 3 fields for validation set A and a single field for validation set B, lesions outside of the fields may remain undetected. Finally, since the data came from screening programs, patients in both datasets were generally referred when moderate-or-worse DR was discovered, and the lack of OCT limited the accuracy of detecting DME. Thus, neither dataset was ideal for evaluation of progression to VTDR.

## Conclusions

Our results suggest that a DLS can be developed to stratify patients by their risk of developing DR. Further studies could help understand the effectiveness of interventions tailored to progression risk (e.g. screening intervals) towards improving health outcomes and reducing cost.

## Data sharing

De-identified data used in this study are not publicly available at present. Interested parties should contact the authors for potential collaborations.

## Acknowledgements

We thank Jacqueline Shreibati, Ellery Wulczyn, and Michael Howell for review and suggestions for the manuscript, and Roy Lee, Noemi Figueroa, and the labeling software team in Google Health for assistance in data labeling.

## References

1. Cheung N, Mitchell P, Wong TY. Diabetic retinopathy. The Lancet. 2010;376(9735):124–36.

2. Ogurtsova K, da Rocha Fernandes JD, Huang Y, Linnenkamp U, Guariguata L, Cho NH, et al. IDF Diabetes Atlas: Global estimates for the prevalence of diabetes for 2015 and 2040. Diabetes Res Clin Pract. 2017 Jun;128:40–50.

3. Scanlon PH. Screening Intervals for Diabetic Retinopathy and Implications for Care. Curr Diab Rep. 2017 Sep 5;17(10):96.

4. Fong DS, Aiello L, Gardner TW, King GL, Blankenship G, Cavallerano JD, et al. Retinopathy in diabetes. Diabetes Care. 2004 Jan;27 Suppl 1:S84–7.

5. Ting DSW, Cheung GCM, Wong TY. Diabetic retinopathy: global prevalence, major risk factors, screening practices and public health challenges: a review. Clin Experiment Ophthalmol. 2016 May;44(4):260–77.

6. Solomon SD, Chew E, Duh EJ, Sobrin L, Sun JK, VanderBeek BL, et al. Diabetic Retinopathy: A Position Statement by the American Diabetes Association. Diabetes Care. 2017;40(3):412–8.

7. International Council of Ophthalmology. ICO Guidelines for Diabetic Eye Care. 2017 [cited 2020 Jun 16]. Available from: http://www.icoph.org/downloads/ICOGuidelinesforDiabeticEyeCare.pdf

8. American Academy Of Ophthalmology. Diabetic Retinopathy Preferred Practice Patterns. 2019 [cited 2020 Jun 16]. Available from: https://www.aao.org/preferred-practice-pattern/diabetic-retinopathy-ppp [Accessed June 16, 2020].

9. Jenkins AJ, Joglekar MV, Hardikar AA, Keech AC, O'Neal DN, Januszewski AS. Biomarkers in Diabetic Retinopathy. Rev Diabet Stud. 2015 Aug 10;12(1-2):159–95.

10. Stratton IM, Kohner EM, Aldington SJ, Turner RC, Holman RR, Manley SE, et al. UKPDS

50: risk factors for incidence and progression of retinopathy in Type II diabetes over 6 years from diagnosis. Diabetologia. 2001 Feb;44(2):156–63.

11. Grading diabetic retinopathy from stereoscopic color fundus photographs--an extension of the modified Airlie House classification. ETDRS report number 10. Early Treatment Diabetic Retinopathy Study Research Group. Ophthalmology. 1991 May;98(5 Suppl):786–806.

12. Wilkinson CP, Ferris FL 3rd, Klein RE, Lee PP, Agardh CD, Davis M, et al. Proposed international clinical diabetic retinopathy and diabetic macular edema disease severity scales. Ophthalmology. 2003 Sep;110(9):1677–82.

13. The Royal College of Ophthalmologists Diabetic Retinopathy Guidelines. [cited 2020 Jun 16]. Available from: The Royal College of Ophthalmologists Diabetic Retinopathy Guidelines. 2013. Available at: https://www.rcophth.ac.uk/wp-content/uploads/2014/12/2013-SCI-301-FINAL-DR-GUIDELINES-DEC-2012-updated-July-2013.pdf [Accessed June 16, 2020].

14. Modjtahedi BS, Theophanous C, Chiu S, Luong TQ, Nguyen N, Fong DS. Two-Year Incidence of Retinal Intervention in Patients With Minimal or No Diabetic Retinopathy on Telemedicine Screening. JAMA Ophthalmology. 2019;137(4):445.

15. Ruamviboonsuk P, Krause J, Chotcomwongse P, Sayres R, Raman R, Widner K, et al. Deep learning versus human graders for classifying diabetic retinopathy severity in a nationwide screening program. npj Digital Medicine. 2019;2(1). Available from: http://dx.doi.org/10.1038/s41746-019-0099-8

16. Grading Diabetic Retinopathy from Stereoscopic Color Fundus Photographs — An Extension of the Modified Airlie House Classification. Ophthalmology. 2020;127(4):S99–119.

17. Varadarajan AV, Bavishi P, Ruamviboonsuk P, Chotcomwongse P, Venugopalan S, Narayanaswamy A, et al. Predicting optical coherence tomography-derived diabetic macular edema grades from fundus photographs using deep learning. Nat Commun. 2020 Jan 8;11(1):130.

18. Szegedy C, Vanhoucke V, Ioffe S, Shlens J, Wojna Z. Rethinking the Inception Architecture for Computer Vision. In: The IEEE Conference on Computer Vision and Pattern Recognition (CVPR). 2016.

19. google-research. google-research/tf-slim. GitHub. [cited 2020 Jun 17]. Available from: https://github.com/google-research/tf-slim

20. Sundararajan M, Taly A, Yan Q. Axiomatic Attribution for Deep Networks. 2017 [cited 2020 Jun 17]. Available from: http://arxiv.org/abs/1703.01365

21. Springenberg JT, Dosovitskiy A, Brox T, Riedmiller M. Striving for Simplicity: The All Convolutional Net. 2014 [cited 2020 Jun 17]. Available from: http://arxiv.org/abs/1412.6806

22. Xu S, Venugopalan S, Sundararajan M. Attribution in Scale and Space. 2020 [cited 2020

Jun 17]. Available from: http://arxiv.org/abs/2004.03383

23. McKinney SM, Sieniek M, Godbole V, Godwin J, Antropova N, Ashrafian H, et al. International evaluation of an AI system for breast cancer screening. Nature. 2020 Jan;577(7788):89–94.

24. Esteva A, Kuprel B, Novoa RA, Ko J, Swetter SM, Blau HM, et al. Dermatologist-level classification of skin cancer with deep neural networks. Nature. 2017 Feb 2;542(7639):115–8.

25. Coudray N, Ocampo PS, Sakellaropoulos T, Narula N, Snuderl M, Fenyö D, et al. Classification and mutation prediction from non–small cell lung cancer histopathology images using deep learning. Nature Medicine. 2018;24(10):1559–67.

26. Krause J, Gulshan V, Rahimy E, Karth P, Widner K, Corrado GS, et al. Grader Variability and the Importance of Reference Standards for Evaluating Machine Learning Models for Diabetic Retinopathy. Ophthalmology. 2018 Aug;125(8):1264–72.

27. Ting DSW, Cheung CY-L, Lim G, Tan GSW, Quang ND, Gan A, et al. Development and Validation of a Deep Learning System for Diabetic Retinopathy and Related Eye Diseases Using Retinal Images From Multiethnic Populations With Diabetes. JAMA. 2017 Dec 12;318(22):2211–23.

28. Gulshan V, Peng L, Coram M, Stumpe MC, Wu D, Narayanaswamy A, et al. Development and Validation of a Deep Learning Algorithm for Detection of Diabetic Retinopathy in Retinal Fundus Photographs. JAMA. 2016 Dec 13;316(22):2402–10.

29. Natarajan S, Jain A, Krishnan R, Rogye A, Sivaprasad S. Diagnostic Accuracy of Community-Based Diabetic Retinopathy Screening With an Offline Artificial Intelligence System on a Smartphone. JAMA Ophthalmol. 2019 Aug 8; Available from: http://dx.doi.org/10.1001/jamaophthalmol.2019.2923

30. Gargeya R, Leng T. Automated Identification of Diabetic Retinopathy Using Deep Learning. Ophthalmology. 2017 Jul;124(7):962–9.

31. Abràmoff MD, Lou Y, Erginay A, Clarida W, Amelon R, Folk JC, et al. Improved Automated Detection of Diabetic Retinopathy on a Publicly Available Dataset Through Integration of Deep Learning. Invest Ophthalmol Vis Sci. 2016 Oct 1;57(13):5200–6.

32. Abràmoff MD, Lavin PT, Birch M, Shah N, Folk JC. Pivotal trial of an autonomous AI-based diagnostic system for detection of diabetic retinopathy in primary care offices. NPJ Digit Med. 2018 Aug 28;1:39.

33. Mitani A, Huang A, Venugopalan S, Corrado GS, Peng L, Webster DR, et al. Detection of anaemia from retinal fundus images via deep learning. Nat Biomed Eng. 2020 Jan;4(1):18–27.

34. Poplin R, Varadarajan AV, Blumer K, Liu Y, McConnell MV, Corrado GS, et al. Prediction of cardiovascular risk factors from retinal fundus photographs via deep learning. Nat Biomed

Eng. 2018 Mar;2(3):158–64.

35. Medeiros FA, Jammal AA, Thompson AC. From Machine to Machine: An OCT-Trained Deep Learning Algorithm for Objective Quantification of Glaucomatous Damage in Fundus Photographs. Ophthalmology. 2019 Apr;126(4):513–21.

36. Phene S, Dunn RC, Hammel N, Liu Y, Krause J, Kitade N, et al. Deep Learning and Glaucoma Specialists: The Relative Importance of Optic Disc Features to Predict Glaucoma Referral in Fundus Photographs. Ophthalmology. 2019 Dec;126(12):1627–39.

37. Arcadu F, Benmansour F, Maunz A, Willis J, Haskova Z, Prunotto M. Deep learning algorithm predicts diabetic retinopathy progression in individual patients. NPJ Digit Med. 2019 Sep 20;2:92.

38. Yan Q, Weeks DE, Xin H, Swaroop A, Chew EY, Huang H, et al. Deep-learning-based Prediction of Late Age-Related Macular Degeneration Progression. Nat Mach Intell. 2020 Feb;2(2):141–50.

39. Bhuiyan A, Wong TY, Ting DSW, Govindaiah A, Souied EH, Theodore Smith R. Artificial Intelligence to Stratify Severity of Age-Related Macular Degeneration (AMD) and Predict Risk of Progression to Late AMD. Translational Vision Science & Technology. 2020;9(2):25.

40. Yim J, Chopra R, Spitz T, Winkens J, Obika A, Kelly C, et al. Predicting conversion to wet age-related macular degeneration using deep learning. Nat Med. 2020 May 18; Available from: http://dx.doi.org/10.1038/s41591-020-0867-7

41. Thakur A, Goldbaum M, Yousefi S. Predicting glaucoma prior to its onset using deep learning. Available from: http://dx.doi.org/10.1101/828681

42. Estil S, Steinarsson AÞ, Einarsson S, Aspelund T, Stefánsson E. Diabetic eye screening with variable screening intervals based on individual risk factors is safe and effective in ophthalmic practice. Acta Ophthalmol. 2020 Mar 25; Available from: https://doi.org/10.1111/aos.14425

43. Cunha-Vaz J, Ribeiro L, Costa M, Simó R. Diabetic Retinopathy Phenotypes of Progression to Macular Edema: Pooled Analysis From Independent Longitudinal Studies of up to 2 Years' Duration. Invest Ophthalmol Vis Sci. 2017 May 1;58(6):BIO206–10.

44. Bearse MA Jr, Adams AJ, Han Y, Schneck ME, Ng J, Bronson-Castain K, et al. A multifocal electroretinogram model predicting the development of diabetic retinopathy. Prog Retin Eye Res. 2006 Sep;25(5):425–48.

45. Rogers SL, Tikellis G, Cheung N, Tapp R, Shaw J, Zimmet PZ, et al. Retinal arteriolar caliber predicts incident retinopathy: the Australian Diabetes, Obesity and Lifestyle (AusDiab) study. Diabetes Care. 2008 Apr;31(4):761–3.

46. Rohan TE, Frost CD, Wald NJ. Prevention of blindness by screening for diabetic retinopathy: a quantitative assessment. BMJ. 1989 Nov 11;299(6709):1198–201.

47. Ferris FL 3rd. How effective are treatments for diabetic retinopathy? JAMA. 1993 Mar

10;269(10):1290–1.

48. Zhao Y, Singh RP. The role of anti-vascular endothelial growth factor (anti-VEGF) in the management of proliferative diabetic retinopathy. Drugs Context. 2018 Aug 13;7:212532.

49. Lin DY, Blumenkranz MS, Brothers RJ, Grosvenor DM. The sensitivity and specificity of single-field nonmydriatic monochromatic digital fundus photography with remote image interpretation for diabetic retinopathy screening: a comparison with ophthalmoscopy and standardized mydriatic color photography. Am J Ophthalmol. 2002 Aug;134(2):204–13.

50. Association AD, American Diabetes Association. 11. Microvascular Complications and Foot Care: Standards of Medical Care in Diabetes−2020. Diabetes Care. 2020;43(Supplement 1):S135–51.

51. Dixon RF, Zisser H, Layne JE, Barleen NA, Miller DP, Moloney DP, et al. A Virtual Type 2 Diabetes Clinic Using Continuous Glucose Monitoring and Endocrinology Visits. J Diabetes Sci Technol. 2019 Nov 25;1932296819888662.

52. Downing J, Bollyky J, Schneider J. Use of a Connected Glucose Meter and Certified Diabetes Educator Coaching to Decrease the Likelihood of Abnormal Blood Glucose Excursions: The Livongo for Diabetes Program. J Med Internet Res. 2017 Jul 11;19(7):e234.

53. tensorflow. tensorflow/models. GitHub. [cited 2020 Jun 17]. Available from: https://github.com/tensorflow/models

54. Kingma DP, Ba J. Adam: A Method for Stochastic Optimization. 2014 [cited 2020 Jun 17]. Available from: http://arxiv.org/abs/1412.6980

# Tables

**Table 1. Baseline characteristics.**

| Characteristics | | Development Dataset | | Validation Datasets | |
|---|---|---|---|---|---|
| | | Training Dataset (EyePACS) | Tuning Dataset (EyePACS) | Validation Dataset A (EyePACS) | Validation Dataset B (Thailand) |
| No. of patients | | 253,598 | 36,228 | 72,457 | 6,791 |
| No. of eyes | | 503,527 | 71,904 | 143,864 | 13,573 |
| No. of eyes with at least one set of images with available grades | | 365,144 | 52,456 | 104,695 | 6,307 |
| No. of eyes selected (random eye per patient in validation sets; see Methods) | | | | 52,403 | |
| No. of eyes with at least two sets of gradable images (on different visits) | | 63,281 | 9,333 | 9,124 | 5,603 |
| No. of eyes with at least two sets of gradable images and without DR at the first visit (No. of patients) | | 55,561 (31,449) | 8,211 (4,651) | 7,976 (7,976) | 4,762 (4,762) |
| Age (years), median (IQR) | | 53 (46-60) | 54 (45-60) | 54 (45-60) | 59 (52-66) |
| No. of females / total* (%) | | 19,919 / 31,447 (63.3%) | 2,905 / 4,649 (62.5%) | 5,118 / 7,975 (64.2%) | 3,277 / 4,759 (68.9%) |
| Race / Ethnicity (%) | | Hispanic: 59.0%<br>White: 7.3%<br>Asian / Pacific Islander: 5.3%<br>Black: 4.0%<br>Native American: 1.3%<br>Other: 1.5%<br>Not available: 21.5% | Hispanic: 58.3%<br>White: 7.4%<br>Asian / Pacific Islander: 4.9%<br>Black: 3.8%<br>Native American: 1.2%<br>Other: 1.6%<br>Not available: 22.7% | Hispanic: 58.7%<br>White: 7.6%<br>Asian / Pacific Islander: 4.8%<br>Black: 4.6%<br>Native American: 1.2%<br>Other: 1.7%<br>Not available: 21.5% | Thai: 100% |
| HbA1c (%), median, (IQR) | | 6.5 (7.3-8.9) | 6.5 (7.2-8.9) | 6.5 (7.3-9.0) | 6.4 (7.2-8.4) |
| Outcome: mild-or-worse ("Mild+") DR within 2 years | Total no. of eyes with known outcome** | 25,211 | 3,688 | 3,678 | 2,345 |
| | No. of eyes with Mild+ outcome (%) | 4,529 (18.0%) | 635 (17.2%) | 685 (18.6%) | 346 (14.8%) |
| Outcome: moderate-or-worse ("Moderate+") DR within 2 years | Total no. of eyes with known outcome** | 24,393 | 3,609 | 3,554 | 2,292 |
| | No. of eyes with Moderate+ outcome (%) | 2,093 (8.6%) | 303 (8.4%) | 317 (8.9%) | 225 (9.8%) |
| Outcome: vision-threatening DR (VTDR)*** within 2 years | Total no. of eyes with known outcome** | 24,104 | 3,602 | 3,500 | 2,219 |
| | No. of eyes with VTDR outcome (%) | 256 (1.1%) | 37 (1.0%) | 43 (1.2%) | 69 (3.1%) |

Abbreviations: DR, Diabetic Retinopathy; IQR, Interquartile Range; HbA1c, Glycated hemoglobin. *A small number of patients did not have this information available. **See "Methods" for binary endpoint definition. ***includes diabetic macular edema.

**Table 2. Predictive performance by the deep learning system (DLS) applied to color fundus photographs (CFPs), in comparison to and in combination with known risk factors for (A) the EyePACS validation dataset and (B) the Thailand dataset.** The highest value(s) in each row are highlighted. A more comprehensive set of risk factors are reported in Supplementary Table S5. *The only risk factor in validation set B that was also available and prognostic in the development set was HbA1c.

**A**

| Risk factor(s) | No. of eyes with risk factor(s) available | No. of eyes that developed DR within 2 years | AUC [95%CI] | | | | |
|---|---|---|---|---|---|---|---|
| | | | Risk factor(s) alone | Using 1-field CFP | | Using 3-field CFPs | |
| | | | | DLS alone | Risk factor(s) + DLS | DLS alone | Risk factor(s) + DLS |
| HbA1c | 2,534 | 458 (18%) | 0.68 [0.65, 0.71] | 0.78 [0.75, 0.80] | 0.79 [0.77, 0.82] | 0.79 [0.76, 0.81] | **0.81 [0.78, 0.83]** |
| Years with diabetes | 3,223 | 615 (19%) | 0.64 [0.61, 0.66] | 0.77 [0.75, 0.79] | 0.78 [0.76, 0.80] | 0.79 [0.76, 0.81] | **0.80 [0.77, 0.82]** |
| Self-reported diabetic control | 1,589 | 274 (17%) | 0.64 [0.61, 0.68] | 0.77 [0.73, 0.80] | 0.78 [0.75, 0.81] | 0.79 [0.76, 0.82] | **0.80 [0.77, 0.84]** |
| Insulin use | 3,678 | 685 (19%) | 0.59 [0.58, 0.61] | 0.78 [0.76, 0.80] | 0.78 [0.76, 0.80] | 0.79 [0.77, 0.81] | **0.80 [0.78, 0.82]** |
| HbA1c, years with diabetes, self-reported diabetic control, insulin use | 1,257 | 221 (18%) | 0.72 [0.68, 0.76] | 0.77 [0.73, 0.81] | 0.79 [0.75, 0.82] | 0.79 [0.75, 0.82] | **0.81 [0.77, 0.84]** |

**B**

| Risk factor(s)* | No. of eyes with risk factor(s) available | No. of eyes that developed DR within 2 years | AUC (95%CI) | | |
|---|---|---|---|---|---|
| | | | Risk factor(s) alone | Using 1-field CFP | |
| | | | | DLS alone | Risk factor(s) + DLS |
| HbA1c | 1,788 | 268 (15%) | 0.62 [0.58, 0.66] | 0.70 [0.67, 0.74] | **0.71 [0.68, 0.75]** |

Abbreviations: DR, Diabetic Retinopathy; AUC, Area Under the Curve; CFP, Color Fundus Photograph; DLS, Deep Learning System; HbA1c, Glycated hemoglobin.

# Figures

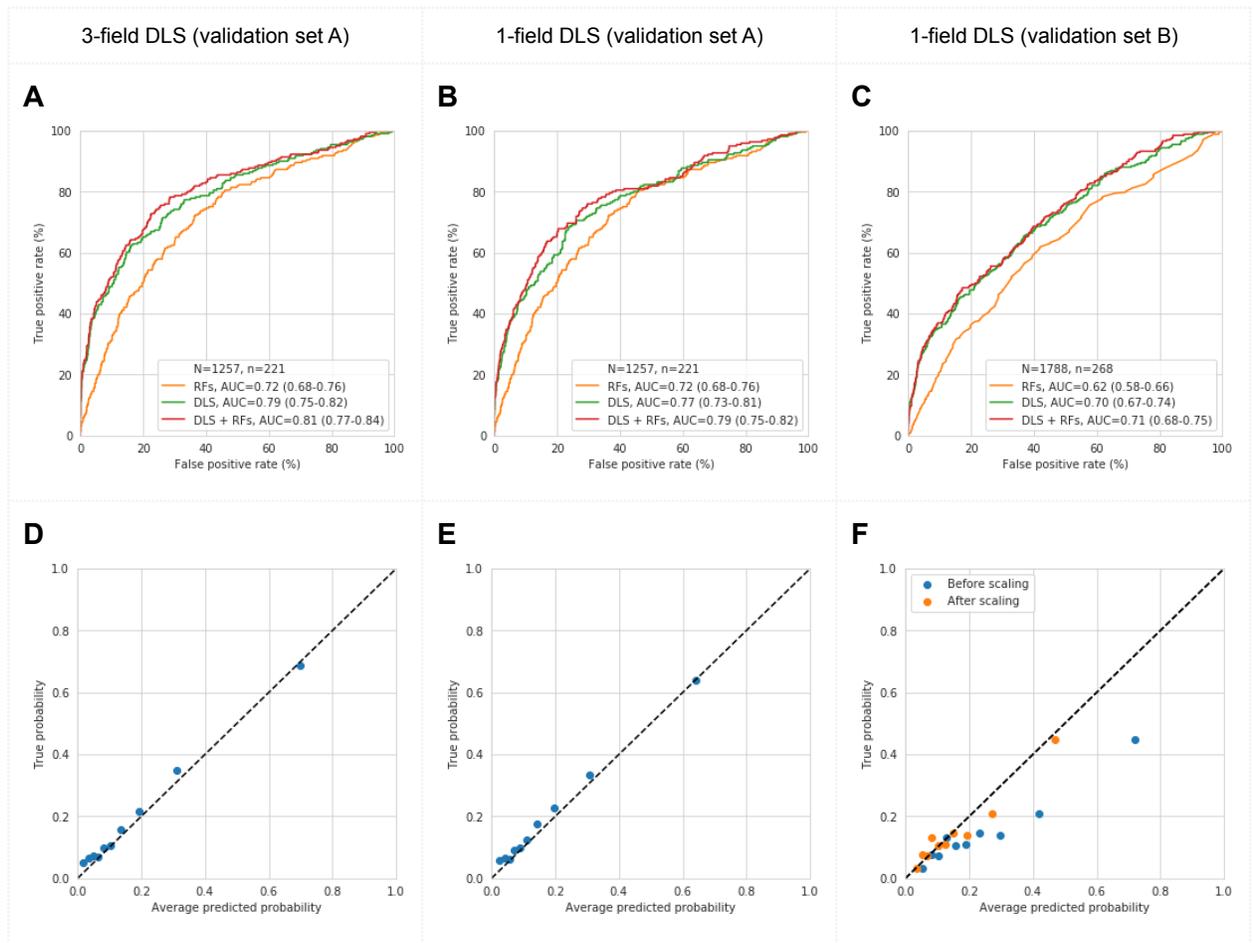

**Figure 1**. Discrimination and calibration of the deep learning model for predicting incidence of diabetic retinopathy (DR). (**A,B**) Receiver operating characteristic (ROC) curves for 3-field and 1-field DLS for validation set A. Curves are shown for risk factors in Table 2, for the DLS alone, and for the combination of these risk factors (RFs) and the DLS. To enable comparison, ROCs were plotted only for patients with known values for the RFs; the ROC for the DLS on the entire validation set is presented in Supplementary Figure S3A-C. (**C**) ROC curve for the 1-field DLS for validation set B. (**D,E,F**) Calibration plots corresponding to the models in panels A-C, for the entire validation datasets. Each plot compares actual vs. average predicted risk, with each blue dot representing a decile of predicted risk. Because the DLS was less well calibrated on validation set B, the DLS was recalibrated and visualized in panel F as orange dots.

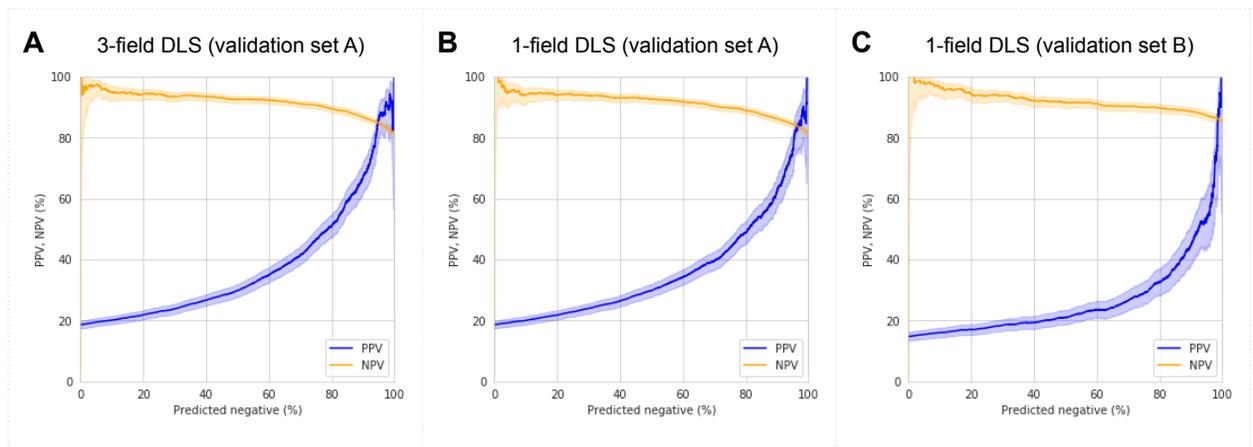

**Figure 2**. Positive predictive value (PPV, blue line) and negative predictive value (NPV, orange line) of the deep learning system (DLS) for predicting development of diabetic retinopathy in (**A,B**) validation set A using 3-field and 1-field models respectively. (**C**) Analysis for validation set B using the 1-field model. The shaded regions indicate 95% confidence intervals.

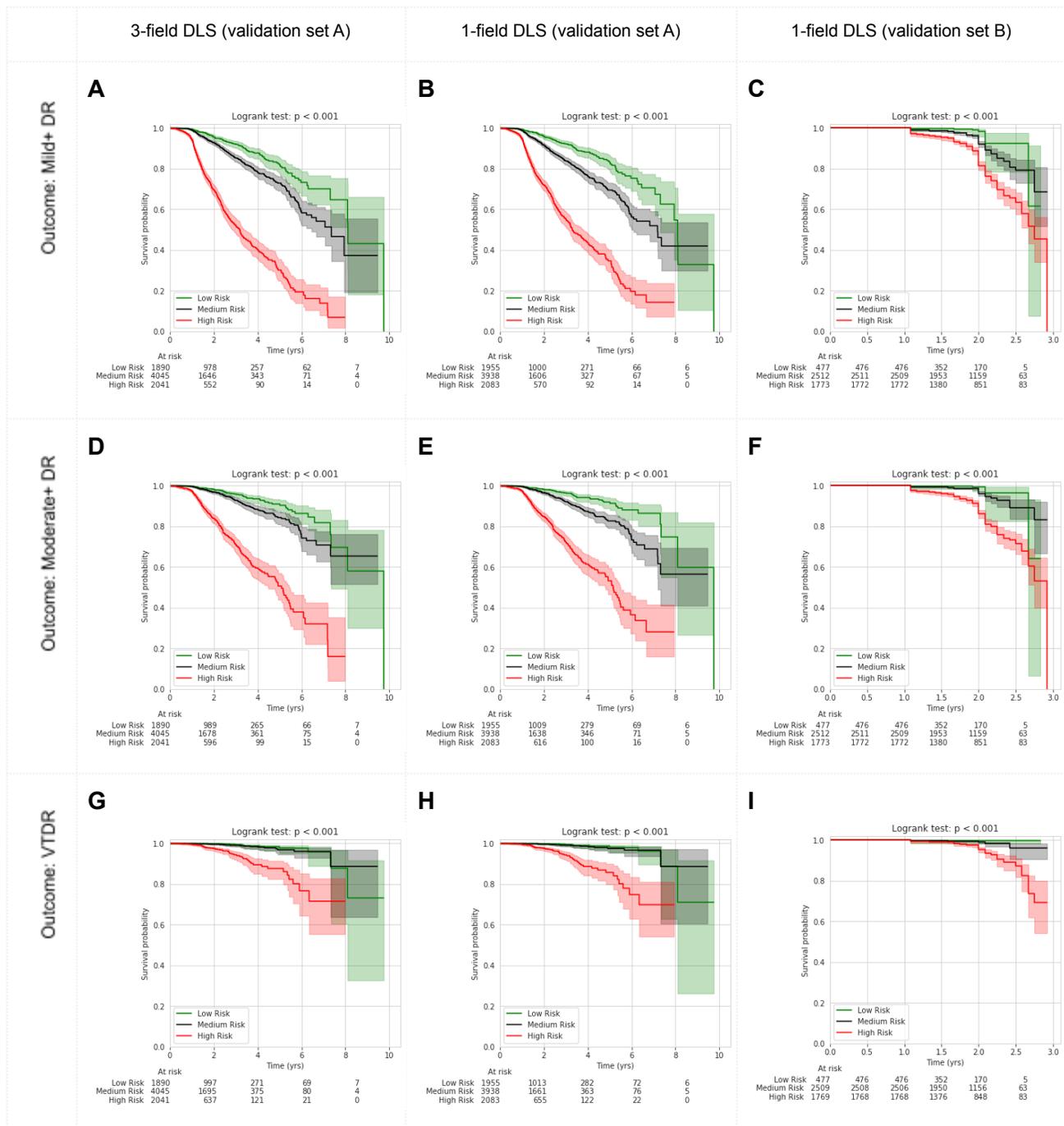

**Figure 3**. Kaplan Meier plot illustrating the incidence of diabetic retinopathy (DR). The red, black, and green lines represent high risk, moderate risk, and low risk as stratified by the deep learning system using the upper and lower quartiles in the tuning dataset. (**A,B**) Plots for 3-field and 1-field DLS for validation set A. (**C**) Plot for 1-field DLS for validation set B. Note that validation sets A and B have different follow-up durations. (**D-F**) represent the same risk groups in panels A-C progressing to moderate-or-worse DR; (**G-I**) represent the same risk groups in panels A-C progressing to vision-threatening DR (VTDR).

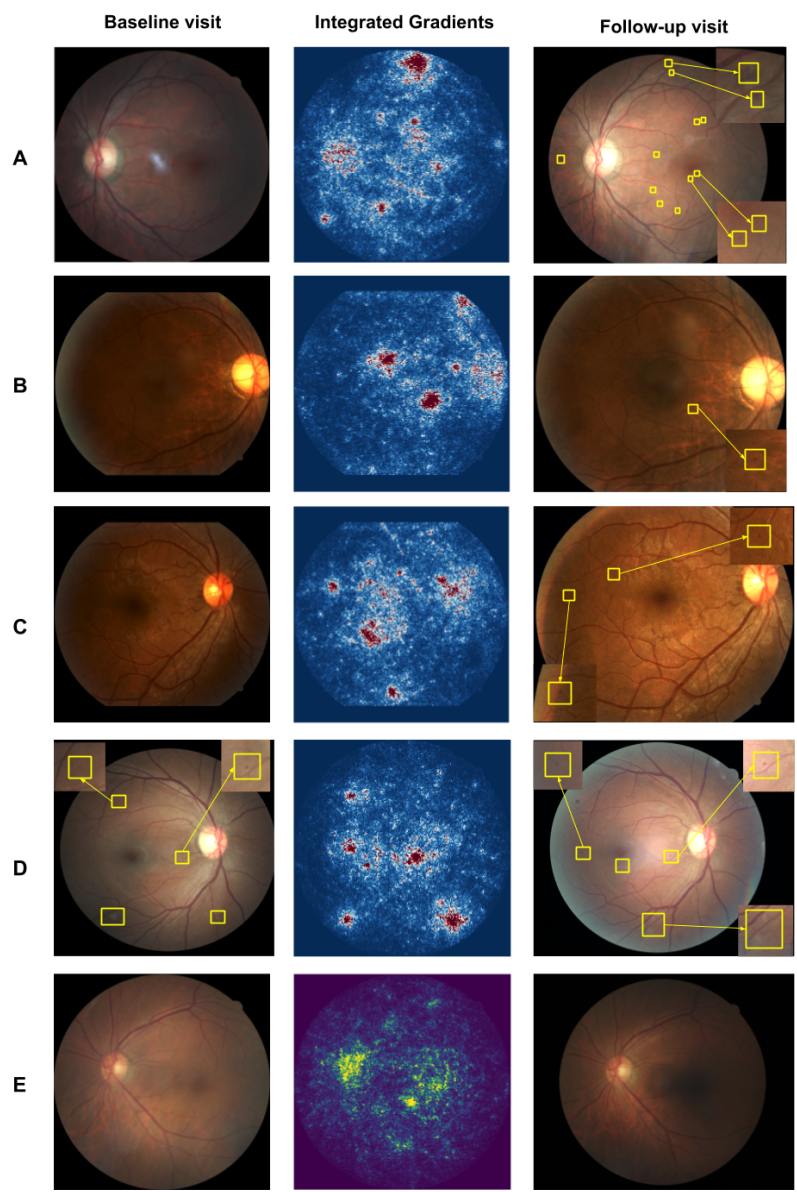

**Figure 4**. Example cases with a color fundus photograph (CFP) at baseline (leftmost column), saliency heatmaps (middle column) using Integrated Gradients[22] explanation technique on the CFP from the baseline visit, and a follow-up visit (rightmost column). In each saliency map, the color represents the strength of the contribution towards the prediction; red indicates a contribution towards developing DR, yellow indicates contribution towards not developing DR, whereas blue indicates little contribution. **(A)** This case did not have diabetic retinopathy (DR) at baseline visit (age=43 years), but developed DR on a future follow-up visit (age=44 years). The prediction was high (0.75) and the highlighted regions in the saliency map eventually developed microaneurysms. Note that not all future microaneurysms were highlighted. (**B**) This case also did not have diabetic retinopathy (DR) at baseline visit (age=48 years), but developed DR on a future follow-up visit (age=50 years). The prediction was high (0.71) and the highlighted regions in the saliency map eventually developed microaneurysms. (**C**) This case did not have DR at the baseline visit (age=34 years), but developed DR at the followup visit (age 37 years). The prediction was high (0.68), but the highlighted regions did not show microaneurysms at the follow-up visit. **(D)** This case had a very high DLS prediction (0.97), and upon closer inspection the baseline image showed subtle signs of mild DR, as evidenced by microaneurysms. The saliency map correctly highlighted the microaneurysms at baseline. **(E)** This case did not have DR at the baseline visit (age=64 years) and also did not develop DR at the latest available follow-up visit (age=70 years). The DLS correctly assigned a small probability of progression (0.06).

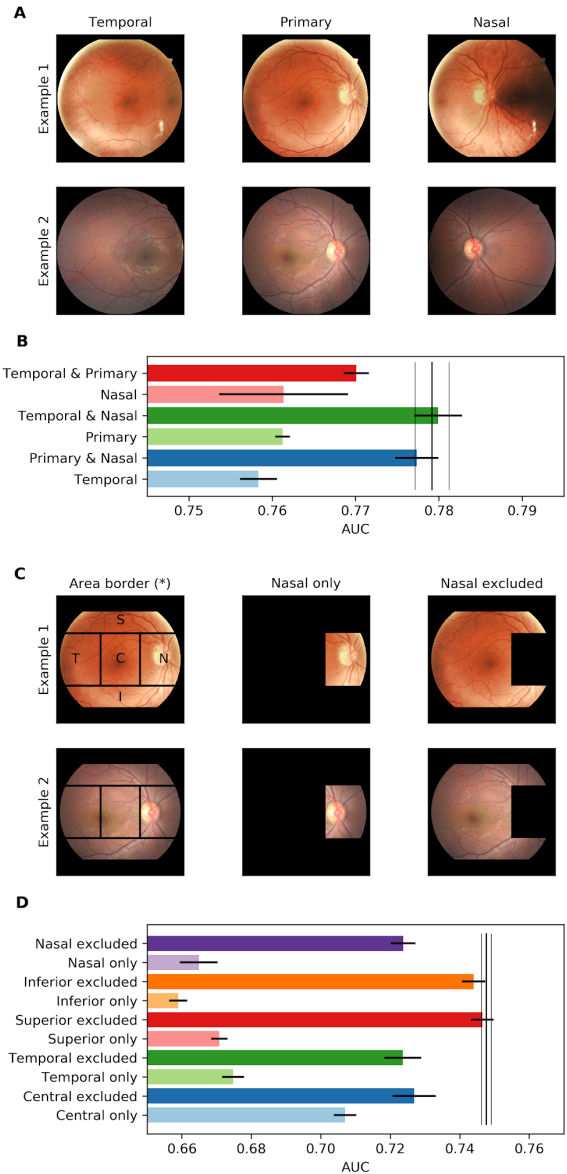

**Figure 5.** Importance of different fields and different parts of the image. **A.** Sample images of each of the three fields. **B.** Deep learning system (DLS) performance using one or two images from three-field images compared to the performance using all three images (vertical black lines). The thick vertical black line indicates the mean, and the thin vertical black lines indicate the standard deviation across three training runs. **C.** The left column contains sample images of the primary field with several different regions indicated by letters: S: Superior, T: Temporal, C: Central, N: Nasal, and I: Inferior. The middle column contains sample images with only one of the regions (Nasal in this example). The right column contains sample images that exclude one region (Nasal in this example). **D.** DLS performance after systematically including only each region or excluding that region, compared to the performance using all the regions (vertical black lines). The thick vertical black line indicates the mean, and the thin vertical black lines indicate the standard deviation across three training runs. Error bars (horizontal black lines) similarly indicate one standard deviation across three training runs.

# Supplement

## Supplementary Methods

### Visualization of differences between patients with vs those without followup

To better understand if there were differences between the patients who had follow-up visits in our dataset vs those who did not, we visualized the differences in several baseline variables as well as the DLS prediction (Supplementary Table S7).

### Handling outlier values for risk factors

To ensure the logistic regression models were not inordinately influenced by outliers in the risk factors (e.g. caused by errors in data entry), some processing was performed on the data. For age, values less than 1 or higher than 122 were removed, whereas values between 90 and 122 were set to 90. HbA1c less than 1% and greater than 18% were removed. "Years with diabetes" values were clipped to be between 1 and 20.

### Comparison of different modeling approaches

In addition to the methods described in the main text, we explored several other neural network modeling approaches and selected the best method based on the results on the tune set (Supplementary Table S4). First, we trained a network to grade for DR on the five-point scale (Supplementary Table S1). The five-point prediction of this network resulted in an AUC of 0.58. However, applying logistic regression to the continuous predicted distribution over the five-point scale resulted in a substantially higher AUC of 0.74. Next, we trained a network to grade for the individual lesions listed in Supplementary Table S1. Applying logistic regression to the binary predicted lesions resulted in an AUC of 0.57. Similarly, applying logistic regression to the continuous predicted distribution of the lesions resulted in a substantially higher AUC of 0.76. Finally, the methods described in the main text, which involves directly training a model to predict development of DR while simultaneously "co-training" on current DR/DME grades, lesions, and various future outcomes (Supplementary Table S4) resulted in the highest AUC of 0.80. When evaluating the combined models containing both the risk factors and 3-field CFPs, the trends were identical, with a 0.04-0.17 increase in AUC compared to using CFPs alone.

We further explored how much training data was necessary for the performance observed (Supplementary Figure S6). Our results suggest that the performance of the DLS had not plateaued and could potentially increase further with more training data.

## Using a reference standard based on automated grading

As noted in the main text, some grading errors are inevitable for subtle findings such as microaneurysms. To further validate our DLS, we used an alternative reference standard based on automated grading. Specifically, a deep learning model was trained and tuned using the development sets (see "Methods") to classify the current stage of DR using 3-fields of CFPs. On the tune set, this model had an AUC of 0.954 for classifying no DR vs Mild+ DR. This model was applied to each set of CFPs in validation set A (See "Methods") to obtain a probability distribution over the five grades and the grade with the maximum probability was treated as the automated DR grade. Next, using the automated grade as the reference standard, the DR status at the first visit was evaluated to decide cohort inclusion for each eye; eyes with no DR at the first visit were included. For each included eye, the reference standard for progression endpoint was then computed (see "Endpoint Definition") using the sequence of automated grades at subsequent visits. We evaluated the 3-field DLS as described in the main text on this new cohort with the new reference standard. The DLS had an AUC of 0.80 (95% CI: 0.77, 0.83). This result further validates the performance of our DLS and provides evidence that the learned risk stratification retains prognostic value against an endpoint defined via an accurate deep learning model for classifying current DR grade.

# Supplementary Tables

**Supplementary Table S1**. Mapping of the graded lesions to corresponding DR levels and DME for the EyePACS dataset. Lesions were graded by EyePACS following the ETDRS protocol, and the mapping was customized to be as consistent as possible with mapping used for our graders (Supplementary Table S2).

| Lesions | Grades |
| --- | --- |
| No lesions | No DR |
| Microaneurysms only | Mild |
| Hemorrhages with or without microaneurysms (HMA) < 2A | Moderate |
| Intraretinal microvascular abnormalities < 8A | |
| Hard exudates | |
| Cotton wool spots* | |
| Focal laser scars | |
| HMA $\geq$ 2A | Severe |
| Venous beading | |
| Intraretinal microvascular abnormalities $\geq$ 8A | |
| Neovascularization at disc or elsewhere | Proliferative DR |
| Fibrovascular proliferation | |
| Preretinal hemorrhage | |
| Vitreous hemorrhage | |
| Panretinal photocoagulation scars | |
| Hard exudates* within 2 disc diameter | DME |

Abbreviations: DR, Diabetic Retinopathy; DME, Diabetic Macular Edema;
* Requires the presence of microaneurysms

**Supplementary Table S2.** Definitions of DR levels and DME for the Thailand dataset based on lesions present. The grades were directly provided by retina specialists (Methods).

| Lesions | Grades |
|---|---|
| No lesions | No DR |
| Microaneurysms only | Mild |
| Hemorrhages* | Moderate |
| Hard exudates* | |
| Cotton wool spots* | |
| Focal laser scars | |
| Greater density/count of hemorrhages | Severe |
| Definite venous beading | |
| Prominent IRMAs | |
| Neovascularization at disc or elsewhere | Proliferative DR |
| Fibrovascular proliferation | |
| Preretinal hemorrhage | |
| Vitreous hemorrhage | |
| Panretinal photocoagulation scars | |
| Hard exudates* within 2 disc diameter | DME |

Abbreviations: DR, Diabetic Retinopathy; IRMA, IntraRetinal Microvascular Abnormality;
* Requires the presence of microaneurysms

**Supplementary Table S3.** Model hyperparameters

| Hyperparameter | Value or Setting |
| --- | --- |
| Preprocessing | Fitting a circle, cropping, and resizing the image (see "resolution" below) |
| Image resolution | 587 x 587 pixels |
| Data augmentation | <ul><li>Vertical and horizontal reflections</li><li>Brightness changes (max delta: 0.115 using tf.image.random_brightness)</li><li>Saturation changes (between 0.560 and 1.275, using tf.image.random_saturation)</li><li>Hue changes (between -0.025 and 0.025, using tf.image.random_hue)</li><li>Contrast changes (between 0.9997 and 1.7705, using tf.image.random_contrast)</li></ul> |
| Neural network backbone | Inception-v3[18] |
| Multi-image reduction | Concatenation of feature maps |
| Dropout keep probability | 0.8 |
| Loss | Cross entropy with weight decay = 4e-5 |
| Co-training heads | Current DR and DME grades, lesions, and future Mild+, Moderate+, Severe+, VTDR, and PDR outcomes at 3, 6, 12 and 24 months |
| Network pre-initialization | From ImageNet-pretrained checkpoint [53] |
| Optimization algorithm and settings | Adam[54]: batch size = 8, learning rate: 0.0003, $\beta_1 = 0.9$, $\beta_2 = 0.999$, $\varepsilon = 0.1$ |
| Stopping criteria | maximum number of training steps = 400,000 with early stopping based on performance on tune set |
| Parameter averaging for inference | Exponential decay with factor of 0.9999 |
| Ensemble size for inference | 10 |

Abbreviations: DR, Diabetic Retinopathy; DME, Diabetic Macular Edema; VTDR, Vision Threatening Diabetic Retinopathy; PDR, Proliferative Diabetic Retinopathy.

**Supplementary Table S4.** Comparison of different modeling approaches.

| Modeling approach | AUC on tune set [95% CI] | |
|---|---|---|
| | 3-field CFPs (N=3,688 eyes, n=635 Mild+) | 3-field CFPs and risk factors (N=1,244 eyes, n=239 Mild+) |
| Using predicted 5-class DR grade (one-hot encoded) | 0.58 [0.57, 0.60] | 0.75 [0.71, 0.79] |
| Using predicted distribution of 5-class DR grade (5 numbers indicating likelihood of each grade) | 0.74 [0.72, 0.77] | 0.81 [0.77, 0.84] |
| Using predicted lesions (one-hot encoded) | 0.57 [0.55, 0.58] | 0.74 [0.70, 0.78] |
| Using predicted lesion probability (one per lesion) | 0.76 [0.74, 0.78] | 0.81 [0.78, 0.84] |
| Using deep learning to predict directly | **0.80 [0.78, 0.82]** | **0.84 [0.81, 0.87]** |

Abbreviations: AUC, Area Under the Curve; CFP, Color Fundus Photograph; DR, Diabetic Retinopathy;

**Supplementary Table S5.** Predictive performance by the deep learning system (DLS) applied to color fundus photographs (CFPs), in comparison to known risk factors for validation set A. Race/ethnicity was also available and coded as Asian/Pacific islander, Black, Hispanic, Native American, White, or Other.

|  |  |  |  | 1-field DLS | | 3-field DLS | |
|---|---|---|---|---|---|---|---|
| **Risk factor / Predictor** | N | n | RF-only | Image-only | RF+Image | Image-only | RF+Image |
| Hypertension (self reported) | 927 | 210 (23%) | 0.54 [0.50, 0.58] | 0.79 [0.75, 0.83] | 0.80 [0.76, 0.83] | 0.80 [0.76, 0.84] | 0.81 [0.78, 0.85] |
| Sex (self reported) | 3677 | 685 (19%) | 0.53 [0.51, 0.55] | 0.78 [0.76, 0.80] | 0.78 [0.76, 0.80] | 0.79 [0.77, 0.81] | 0.79 [0.77, 0.81] |
| Aspirin use | 1528 | 271 (18%) | 0.53 [0.50, 0.56] | 0.77 [0.74, 0.80] | 0.77 [0.74, 0.80] | 0.79 [0.75, 0.82] | 0.79 [0.76, 0.82] |
| Sulfonylurea use | 1528 | 271 (18%) | 0.53 [0.50, 0.56] | 0.77 [0.74, 0.80] | 0.77 [0.74, 0.80] | 0.79 [0.75, 0.82] | 0.79 [0.76, 0.82] |
| Ace inhibitor use | 1528 | 271 (18%) | 0.52 [0.49, 0.55] | 0.77 [0.74, 0.80] | 0.77 [0.74, 0.80] | 0.79 [0.75, 0.82] | 0.79 [0.75, 0.82] |
| Statin use | 1528 | 271 (18%) | 0.52 [0.49, 0.55] | 0.77 [0.74, 0.80] | 0.77 [0.74, 0.80] | 0.79 [0.75, 0.82] | 0.79 [0.75, 0.82] |
| Triglycerides level | 1672 | 279 (17%) | 0.52 [0.48, 0.56] | 0.74 [0.70, 0.77] | 0.74 [0.70, 0.77] | 0.76 [0.72, 0.79] | 0.76 [0.73, 0.79] |
| Beta blocker use | 1528 | 271 (18%) | 0.51 [0.50, 0.52] | 0.77 [0.74, 0.80] | 0.77 [0.74, 0.80] | 0.79 [0.75, 0.82] | 0.79 [0.75, 0.82] |
| Age | 3678 | 685 (19%) | 0.51 [0.49, 0.53] | 0.78 [0.76, 0.80] | 0.78 [0.76, 0.80] | 0.79 [0.77, 0.81] | 0.79 [0.77, 0.81] |
| Race / Ethnicity | 2834 | 550 (19%) | 0.51 [0.49, 0.53] | 0.77 [0.75, 0.80] | 0.78 [0.75, 0.80] | 0.79 [0.77, 0.81] | 0.79 [0.77, 0.81] |
| Total cholesterol level | 1687 | 279 (17%) | 0.51 [0.47, 0.55] | 0.74 [0.71, 0.77] | 0.74 [0.71, 0.78] | 0.76 [0.73, 0.79] | 0.76 [0.73, 0.79] |
| Alpha-glucosidase inhibitor use | 1528 | 271 (18%) | 0.50 [0.50, 0.50] | 0.77 [0.74, 0.80] | 0.77 [0.74, 0.80] | 0.79 [0.75, 0.82] | 0.79 [0.75, 0.82] |
| Angiotensin II receptor blocker use | 1528 | 271 (18%) | 0.50 [0.50, 0.50] | 0.77 [0.74, 0.80] | 0.77 [0.74, 0.80] | 0.79 [0.75, 0.82] | 0.79 [0.75, 0.82] |
| Prenatal vitamins use | 1528 | 271 (18%) | 0.50 [0.50, 0.50] | 0.77 [0.74, 0.80] | 0.77 [0.74, 0.80] | 0.79 [0.75, 0.82] | 0.79 [0.75, 0.82] |
| Dipeptidyl peptidase-4 inhibitor use | 1528 | 271 (18%) | 0.50 [0.49, 0.51] | 0.77 [0.74, 0.80] | 0.77 [0.73, 0.80] | 0.79 [0.75, 0.82] | 0.79 [0.75, 0.82] |
| Thiazolidinedione use | 1528 | 271 (18%) | 0.50 [0.48, 0.51] | 0.77 [0.74, 0.80] | 0.77 [0.74, 0.80] | 0.79 [0.75, 0.82] | 0.79 [0.76, 0.82] |
| Biguanide use | 1528 | 271 (18%) | 0.48 [0.45, 0.50] | 0.77 [0.74, 0.80] | 0.77 [0.74, 0.80] | 0.79 [0.75, 0.82] | 0.79 [0.75, 0.82] |

Abbreviations: DLS, Deep Learning System; RF, Risk Factor;

**Supplementary Table S6.** Univariable and multivariable Cox analysis of the DLS in comparisons with risk factors. (**A**) Analysis for validation dataset A; n=2,786 (the number of eyes with all risk factors available). (**B**) Analysis for validation dataset B; n=3,404.

**A**

| Risk factor(s) or predictor | Univariable analysis | | Multivariable analysis | |
|---|---|---|---|---|
| | Hazard ratio [95%CI] | p | Hazard ratio [95%CI] | p |
| DLS (per standard deviation increase) | 1.89 [1.83, 1.97] | <0.001 | 1.80 [1.67, 1.94] | <0.001 |
| HbA1c (per % point increase) | 1.22 [1.19, 1.25] | <0.001 | 1.15 [1.07, 1.22] | <0.001 |
| Self-reported diabetic control: poor | Ref. | N/A | Ref. | N/A |
| Self-reported diabetic control: fair | 0.65 [0.49, 0.86] | 0.003 | 0.91 [0.64, 1.30] | 0.612 |
| Self-reported diabetic control: moderate | 0.62 [0.49, 0.79] | <0.001 | 1.07 [0.77, 1.48] | 0.707 |
| Self-reported diabetic control: good | 0.35 [0.26, 0.46] | <0.001 | 0.62 [0.40, 0.98] | 0.039 |
| Self-reported diabetic control: excellent | 0.26 [0.16, 0.41] | <0.001 | 0.73 [0.41, 1.29] | 0.277 |
| Years with diabetes | 1.08 [1.07, 1.10] | <0.001 | 1.05 [1.02, 1.07] | <0.001 |
| Insulin use | 2.39 [2.10, 2.73] | <0.001 | 1.08 [0.83, 1.39] | 0.561 |

Abbreviations: DLS, Deep Learning System;

**B**

| Risk factor(s) or predictor | Univariable analysis | | Multivariable analysis | |
|---|---|---|---|---|
| | Hazard ratio [95%CI] | p | Hazard ratio [95%CI] | p |
| DLS (per standard deviation increase) | 1.75 [1.64, 1.87] | <0.001 | 1.71 [1.57, 1.86] | <0.001 |
| HbA1c (per % point) | 1.17 [1.11, 1.24] | <0.001 | 1.11 [1.05, 1.19] | <0.001 |
| Self-reported hypertension | 1.11 [0.85, 1.44] | 0.445 | 1.11 [0.82, 1.49] | 0.494 |

Abbreviations: DLS, Deep Learning System; HbA1c, Glycated hemoglobin;

# Supplementary Figures

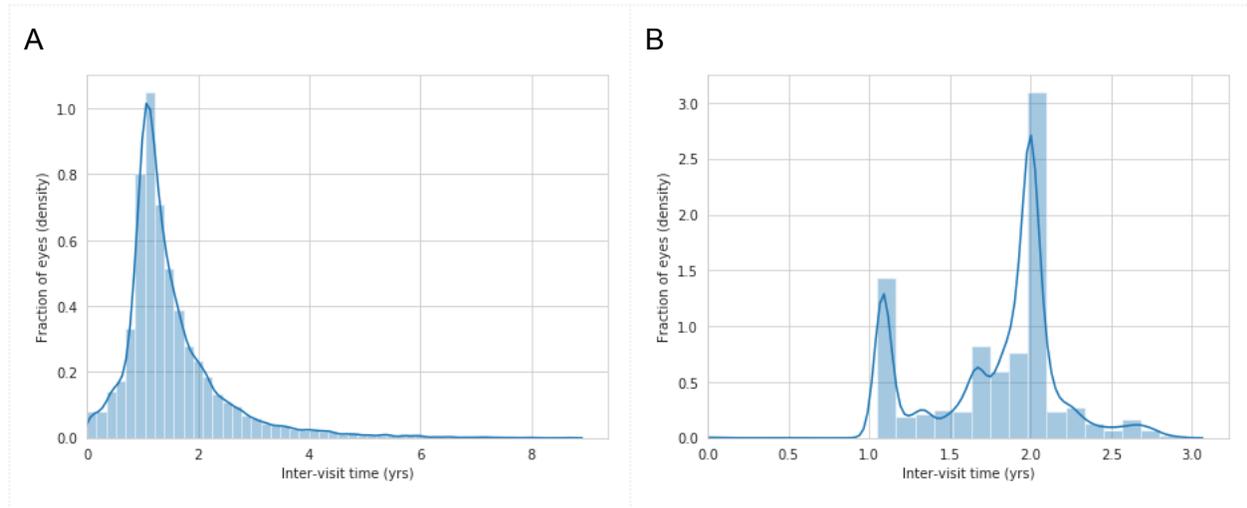

**Supplementary Figure S1**. Histogram of the time in years between visits for (**A**) validation set A and (**B**) validation set B.

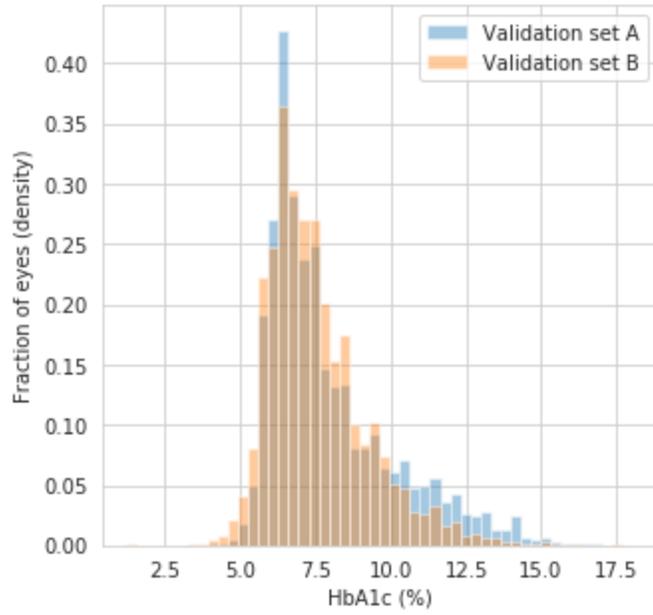

**Supplementary Figure S2.** HbA1c distribution in the two validation datasets.

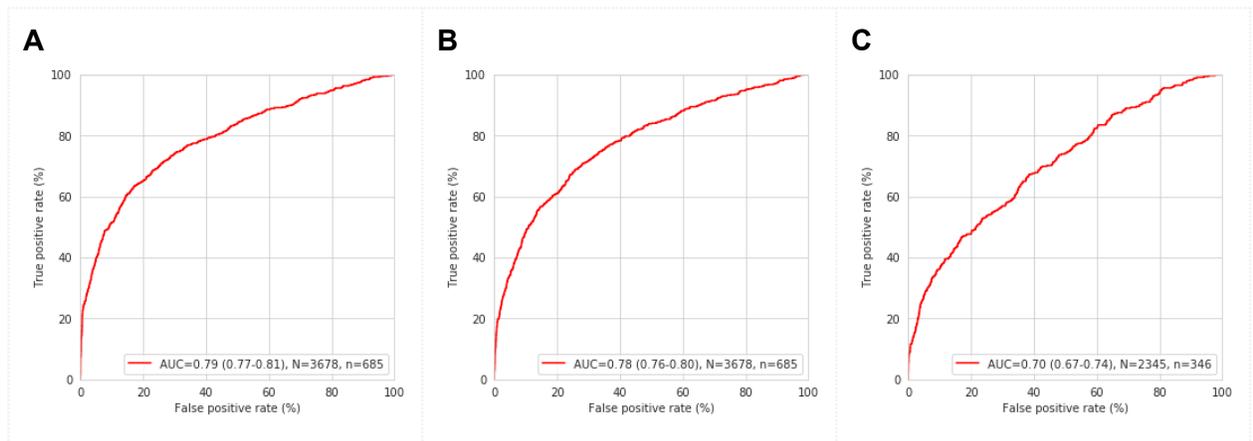

**Supplementary Figure S3**. Discrimination of the deep learning model for predicting incidence of diabetic retinopathy (DR). This is the equivalent of Figure 1, but for all patients instead of only those with the corresponding risk factors. (**A,B**) Receiver operating characteristic (ROC) curves for 3-field and 1-field DLS for validation set A. (**C**) ROC curve for the 1-field DLS for validation set B.

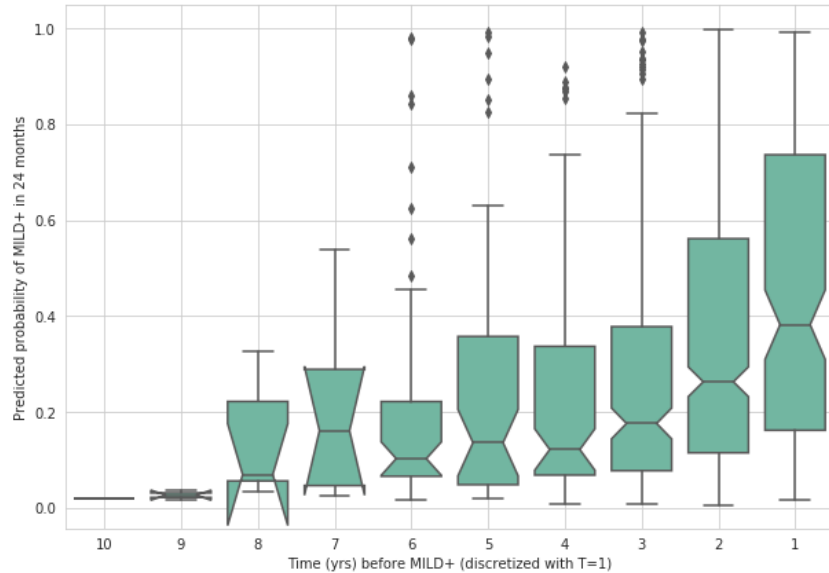

**Supplementary Figure S4.** Predicted probability of developing DR in eyes that eventually developed DR, as a function of time before DR developed. The horizontal lines in the boxes represent the median and quartiles; the whiskers represent 1.5 times the interquartile range; diamonds represent outlier points outside of the whiskers. The notches in the box plot represent confidence intervals across the median, computed using a Gaussian-based asymptotic approximation, as implemented in Python's matplotlib library. Note that only 17 eyes had images 7 years or longer before eventually developing DR, consistent with larger confidence intervals for the median towards the left of the plot.

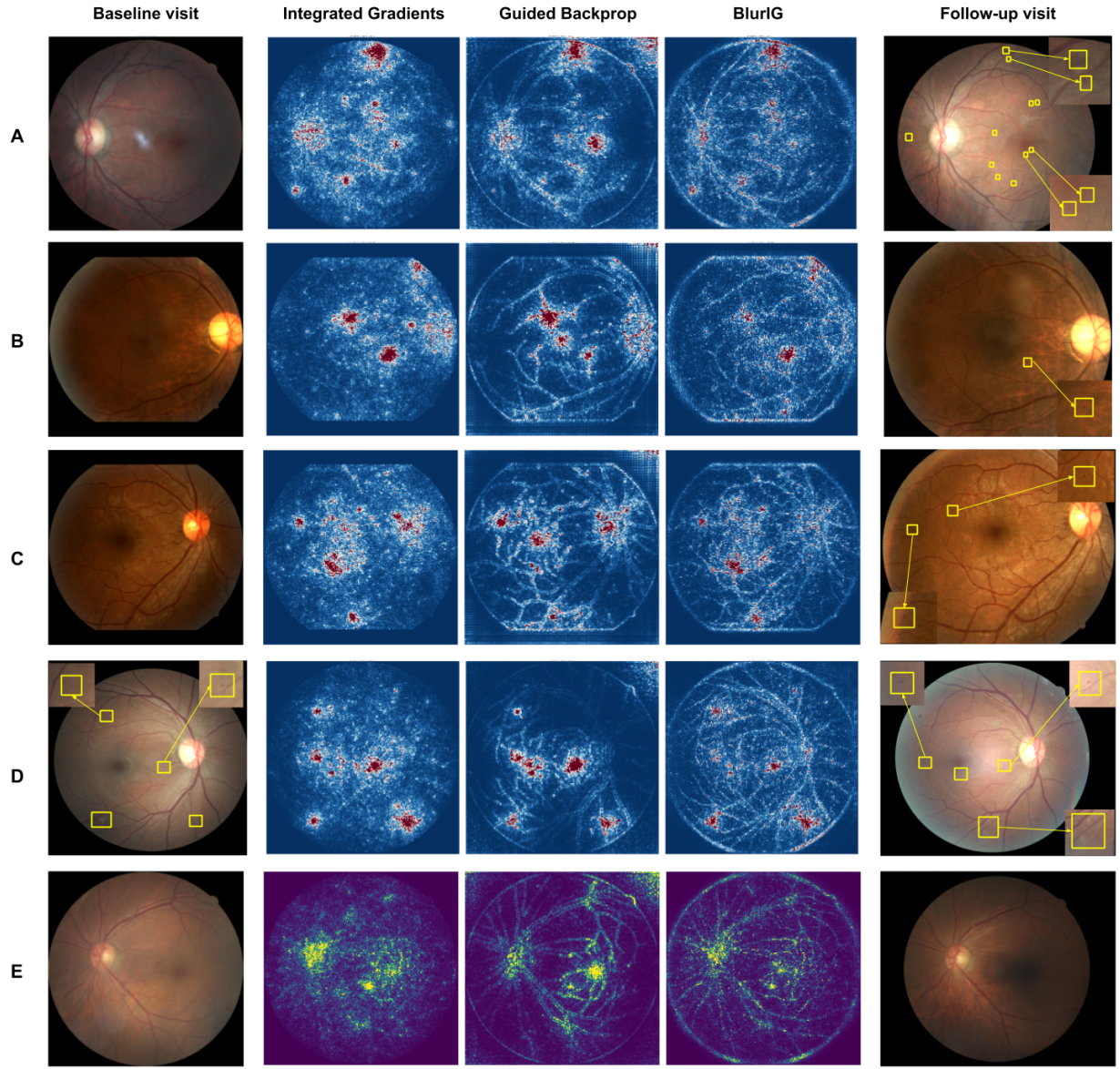

**Supplementary Figure S5.** Example cases presented in Figure 4, but with three different saliency techniques instead of one.

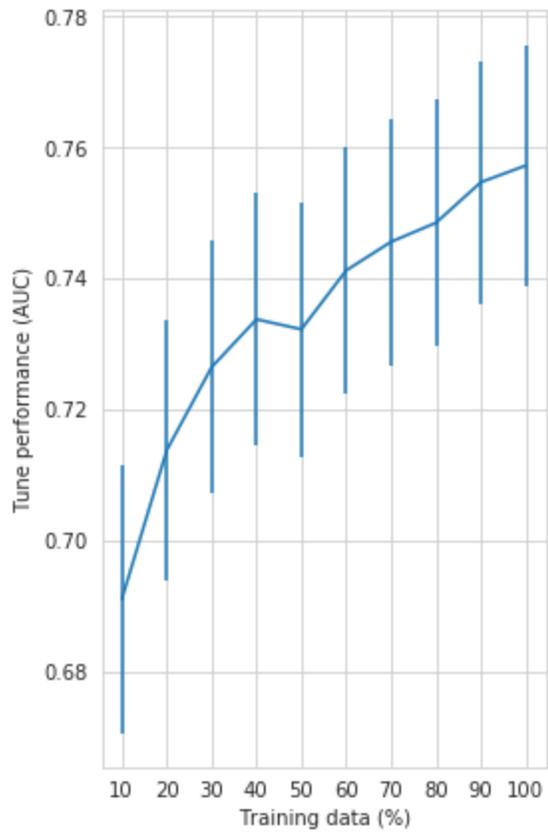

**Supplementary Figure S6**. Effect of amount of training data on the performance of the deep learning system (1-field). Vertical bars indicate 95% confidence intervals across the AUC as computed on the tune set with the DeLong method, from one trained model per point.

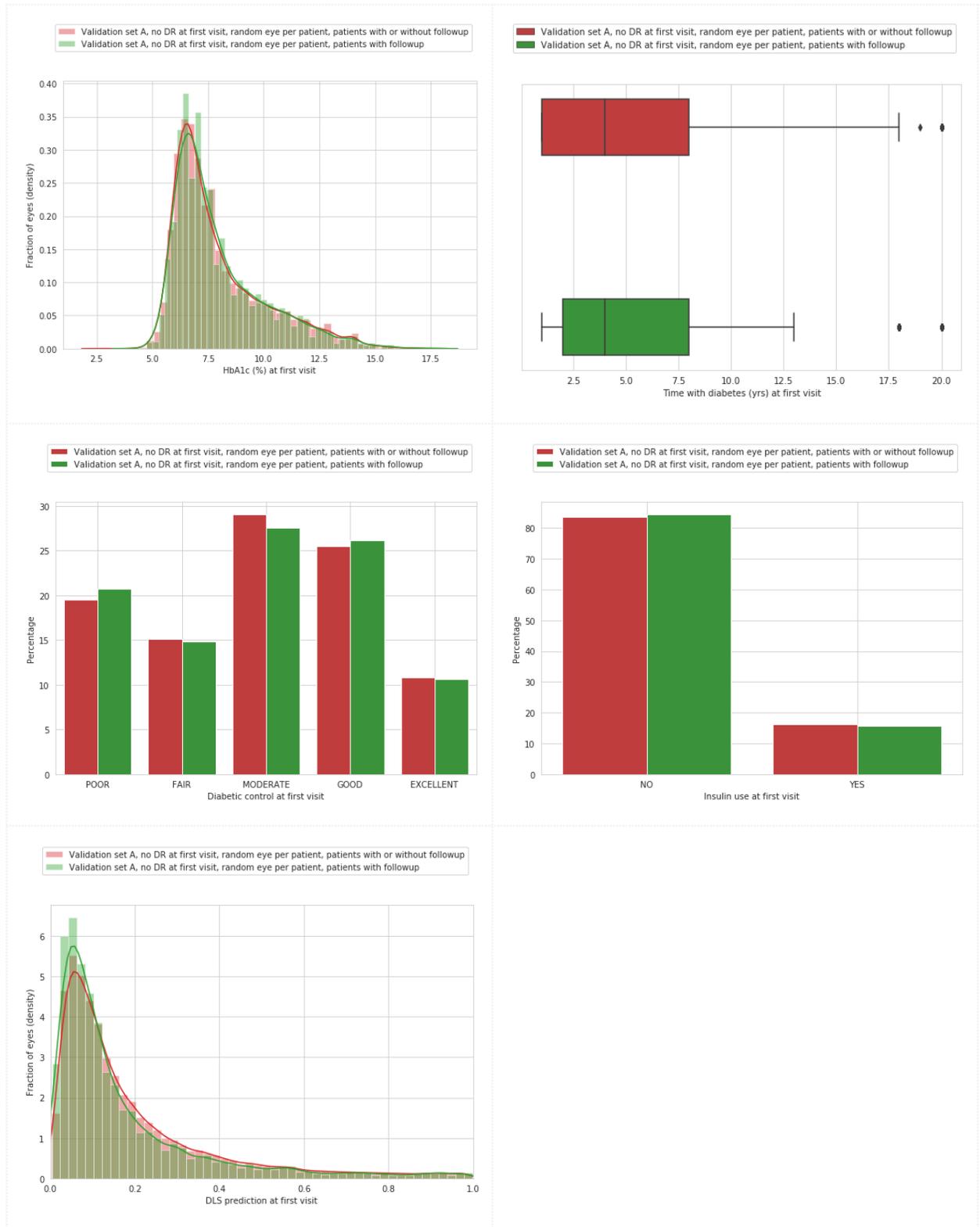

**Supplementary Figure S7.** Differences in baseline variables and the DLS predictions between patients with vs those without followup.